\documentclass[aps,pre,reprint,superscriptaddress,nofootinbib]{revtex4-1}

\usepackage{ amssymb }
\usepackage{graphicx}
\usepackage{float}
\usepackage{subfigure}
\usepackage{booktabs}
\usepackage{array}
\usepackage{natbib}
\usepackage{collref}
\usepackage{tipa}
\usepackage[justification=raggedright,singlelinecheck=false]{caption}
\usepackage[export]{adjustbox}
\newcommand{\squeezeup}{\vspace{-1.25mm}}
\newcommand{\spaceup}{\vspace{+1.25mm}}
\begin{document}


\title{\squeezeup\squeezeup\squeezeup Compensatory interactions to stabilize multiple steady states or mitigate the effects of multiple deregulations in biological  networks\squeezeup}



\author{Gang Yang}
\email{gzy105@psu.edu}

\affiliation{Department of Physics, Pennsylvania State University, University Park, Pennsylvania 16802, USA}

\author{Colin Campbell}
\affiliation{Department of Physics, Pennsylvania State University, University Park, Pennsylvania 16802, USA}
\affiliation{Department of Physics, Washington College, Chestertown, Maryland 21620, USA}

\author{R\'{e}ka Albert}
\email{rza1@psu.edu\squeezeup\squeezeup\squeezeup\squeezeup}

\affiliation{Department of Physics, Pennsylvania State University, University Park, Pennsylvania 16802, USA}


\date{\today}

\begin{abstract}
Complex diseases can be modeled as damage to intra-cellular networks that results in abnormal cell behaviors. Network-based dynamic models such as Boolean models have been employed to model a variety of biological systems including those corresponding to disease. Previous work designed compensatory interactions to stabilize an attractor of a Boolean network after single node damage. We generalize this method to a multi-node damage scenario and to the simultaneous stabilization of multiple steady state attractors. We classify the emergent situations, with a special focus on combinatorial effects, and characterize each class through simulation. We explore how the structural and functional properties of the network affect its resilience and its possible repair scenarios. We demonstrate the method$'$s applicability to two intra-cellular network models relevant to cancer. This work has implications in designing prevention strategies for complex disease.
\end{abstract}


\maketitle

\section{Introduction}
\squeezeup
Complex networks are increasingly used to understand and simulate the behavior of biological systems such as cellular signaling networks \cite{ref2,ref3,ref4,ref5,ref6,ref7}. The network-based dynamic modelling approach aims to capture the biological function and behavior of these systems as an emergent property that arises from the totality of interactions among the components \cite{ref8}. Several researchers have successfully used network-based approaches such as Boolean and logical models to study specific biological processes \cite{ref3,ref4,ref9}. Complex diseases including diabetes and cancers can be modeled as network damage due to temporary or permanent node perturbation (e.g. constitutive activation of a protein arising from a genetic mutation) \cite{ref3, ref10}. Thus the topics of network repair and network control have drawn significant attention in the scientific community \cite{ref1, ref11, ref12,ref13}. Most approaches aim to influence network dynamics by controlling the states of certain nodes of the network \cite{ref11, ref12}. Recently, another approach to the network control problem, namely modifying the interactions in the network, was proposed \cite{ref1}. Using this approach, compensatory interventions can be found to stabilize an attractor (e.g. steady state) of the network after damage to a single node \cite{ref1}. These interventions can be implemented as preventive measures or applied immediately after the onset of damage. The effect of the intervention is that the perturbation does not propagate to the rest of the network, and a close-to-normal behavior is restored \cite{ref1}. Ultimately, a combination of node-based and edge-based approaches will provide researchers more potential therapeutic strategies.

Recent research suggests that complex diseases such as cancer often involve multiple gene mutations and the ``one disease, one target, one drug'' approach may not be effective to battle these diseases \cite{ref10, ref14, ref15}. Thus it is worthwhile to use the network paradigm to explore the combinatorial effect of multiple gene mutations, and to design control measures to prevent these effects. Moreover, many biological systems were shown to have several possible steady states (e.g. several possible cell types), each reachable for alternative histories (time courses) \cite{ref3, ref4, ref8}. Repair interventions should be cognizant of these alternative states and maintain (or eliminate) them as necessary or desired in the specific context. Here we generalize the method of Campbell \textit{et al.} \cite{ref1} to a multiple node damage setting, and to systems that have multiple steady states, aiming to provide a theoretical platform to mitigate damage more realistically. 

This paper presents three key results. First, we use analytical and computational methods to study how network structure and regulatory logic affect the resilience of the network$'$s steady states to single node perturbation. Second, we present an algorithm to design compensatory interventions to stabilize a steady state of the network after double node damage and evaluate it on random Boolean networks. Third, we apply the algorithm on stabilizing two steady states simultaneously after a single node damage and discuss the emerging situations and their corresponding frequencies. We apply the above algorithms to two biological examples and also adapt the latter algorithm to the alternative goal of stabilizing a steady state and destabilizing another. 

\squeezeup\squeezeup
\section{Background: Boolean modeling}
\squeezeup
A network is a mathematical abstraction of a set of relationships between various elements. The network consists of nodes that represent the different elements and edges that specify the pairwise relationships between them \cite{ref16, ref17}. In biological networks at the molecular level, nodes are molecular species such as small molecules, RNA, protein, and edges indicate interactions and regulatory relationships \cite{ref8, ref18}. A substantial amount of studies have characterized the topological properties of networks, such as degree distribution, heterogeneity and community structure \cite{ref16, ref17}. Biological networks were found to exhibit interesting topological properties such as a heterogeneous degree distribution \cite{ref19, ref20}. However, in order to understand the biological function of a system, the network$'$s topological information alone is not enough and dynamical information should be incorporated. More specifically, in a dynamical model, each node $\mathit{i}$ is characterized by a state variable $\mathit{\sigma_{i}}$, which can be continuous or discrete, and the vector $\mathit{(\sigma_{1},\cdots,\sigma_{n})}$ represents the state of the system \cite{ref8}. The state of the system can be followed in continuous time or at discrete time intervals. In discrete time models, the activity of each node $\mathit{\sigma_{i}}$ is described by a regulatory rule $\mathit{\sigma_{i} (t+\tau_{i})=f(\sigma_{i_{1}} (t),\cdots,\sigma_{i_{k}} (t))}$, where $\mathit{i_{1},\cdots,i_{k}}$ are the regulating nodes of $\mathit{i}$ and $\mathit{\tau_{i}}$ is a discrete time delay.

Here we focus on discrete time Boolean network models, where node states are binary, 1(ON) or 0(OFF), and the regulatory rule is specified by a truth table. This is motivated by the fact that biological species are frequently observed to have highly nonlinear regulation and switch-like behavior; thus the node state 1 means the molecular species is above a threshold concentration or activity and the node state 0 means it is below a threshold concentration or activity \cite{ref8, ref18}. The time trajectory of the system is simulated deterministically or stochastically depending on the updating scheme. A simple deterministic updating scheme is synchronous updating, where $\mathit{\tau_{i}=1}$ for every node. For this scheme, given a specific initial state, the system will deterministically evolve into an attractor, which can be a steady state (fixed point) or several states that repeat regularly, called a limit cycle. Steady states can be interpreted as cell types and limit cycles correspond to a cell cycle or circadian rhythms \cite{ref8}. A commonly used stochastic updating scheme is general asynchronous updating, where a random node is selected to be updated at each time step \cite{ref21}. This type of update is motivated by the fact that different biological processes have various timescales, and often the timescales of specific processes are not known \cite{ref22}. Fixed points (steady states) do not depend on the updating scheme \cite{Attractor}. However, limit cycles are generally unstable to infinitesimal deviation from synchronous updating \cite{Attractor}, and the known variety of time scales in biological processes makes limit cycles observed in synchronous updating schemes inherently suspect. While stochastic update may lead to attractors that involve irregular repetitions of a set of states (so-called complex attractors) \cite{ref8}, we here focus our attention on steady state attractors.  Abnormal behavior of a certain element can be modelled as a change in the node state, either a temporary perturbation or permanent damage \cite{ref3, ref10}. For example, a loss-of-function mutation or the knockout of a gene can be represented as a permanent OFF state of the corresponding node in the network.

\squeezeup\squeezeup\squeezeup\squeezeup
\section{Results}
\squeezeup\squeezeup\squeezeup
\subsection{The influence of single node damage on a steady state of a system}
\squeezeup\squeezeup
\label{sec:III.A}
We consider a Boolean model of a biological system; this model will have one or several   attractors. We start from a steady state $\mathit{s}$. Then we consider damage to a node $\mathit{i}$ by permanent knockout (sustained OFF state) or constitutive expression or activity (sustained ON state). If the damaged state $\mathit{s^{*}}$ is a new steady state (i.e. other nodes are not affected by the perturbation), we say that steady state $\mathit{s}$  is stable against the damage. In the converse case, the state of one or more nodes will change, which then has a cascading effect in the biological system. We say the steady state $\mathit{s}$ needs repair in order to prevent damage propagation. We define the sensitive node set $\mathit{S_{i}}$ as the set of nodes that would change their state as a direct consequence of the damage to node $\mathit{i}$. 

Previous research has studied the relationship between a network$'$s structure and its topological resilience to incremental node loss  \cite{ref23} and the relationship between average degree and the effect of single node damage \cite{ref1}. It was shown that the larger the average node degree, the less stable a steady state is against single node damage. Another related result is that  random Boolean network ensembles will go through a phase transition from a frozen phase to a chaotic phase as the average node degree increases. Two states that initially differ in a single node's state will diverge on average in the chaotic phase. The critical boundary is average degree $\mathrm{<}K\mathrm{>}=2$ when considering unbiased Boolean logic (all Boolean functions) and using an annealed approximation (at every time step the input nodes and Boolean functions are randomized for each node)\cite{ref24,ref30,ref29,ref28,semiannealed,PhysRevE55257}. We note that our setting of a steady state damaged by a single node knockout is different from what was considered in previous work on random Boolean network ensembles. 

As biological networks have been observed to exhibit degree heterogeneity and long-tailed decreasing degree distributions \cite{ref2, ref19, ref25, ref26}, we explore the effect of degree heterogeneity on the resilience of a steady state following single node knockout. To probe a variety of regulatory rules consistent with a given number of regulators, we first consider random Boolean rules, and then focus on more realistic nested canalizing Boolean functions. 

\squeezeup\squeezeup\squeezeup
\subsubsection{Theoretical estimation of resilience probability in case of single node damage}
\squeezeup\squeezeup
\label{sec:III.A.1}
We define the resilience probability $\mathit{(RP)}$ of a steady state as the probability that the steady state of the network is stable against single node damage. It follows that the damage probability $\mathit{DP=1-RP}$. We define $\mathit{\alpha(I_{j} )}$ to be the probability that a node $\mathit{j}$ with in-degree $\mathit{I_{j}}$ is stable (does not change state) if one of its randomly chosen inputs, $\mathit{i}$, is knocked out. Knocking out a node $\mathit{i}$ will directly affect the state of at most $\mathit{O_{i}}$ nodes, where $\mathit{O_{i}}$ is the out-degree of node $\mathit{i}$. If we denote the nodes regulated by $\mathit{i}$ as $\mathit{n_{1},n_{2},\cdots,n_{O_{i}}}$, then the probability that the state of the system is a steady state after we knock out node $\mathit{i}$ alone is $\mathit{p(i)=\prod_{i=1}^{O_{i}}\alpha(I_{n_i })}$  since every regulated node must be stable for the overall network to be stable. The average $\mathit{RP}$ is given by $\mathit{RP=\frac{1}{N} \sum_{i=1}^{N}p(i)}$. Under the mean-field assumption that every node follows the same node in-degree distribution   $\mathit{f(I_{i})}$ and out-degree distribution $\mathit{g(O_{i})}$, the average $\mathit{RP}$ can be estimated as\footnote{To be exact, the in-degree distribution  $\mathit{f(I_{i})}$ in formula \ref{equ:RP} should be the conditional in-degree distribution conditioned on a node being knocked out. The conditional in-degree distribution can be obtained through the in-degree distribution reweighted by in-degree.}

\squeezeup\squeezeup\squeezeup\squeezeup
\begin{equation}
\label{equ:RP}
RP=\sum_{O_{i}} g(O_{i})(\sum_{I_{j}}f(I_{j})\alpha(I_{j}))^{O_{i}}
\squeezeup\squeezeup
\end{equation}

In all cases $\mathit{\alpha(I_{j}=0)=1}$ as a source node cannot, by definition, be disrupted by any other node. If each possible Boolean function occurs with equal chance, $\mathit{\alpha(I_{j} )=\frac{1}{2}}$ for $\mathit{I_{j}>0}$; this is due to the equal probability of having 0 or 1 values in each position of the function, which leads to a chance of one half that a change in value of an input variable does not lead to a change in value of the output\footnote{A similar result was obtained in \cite{Kauffman2}. \squeezeup\squeezeup\squeezeup\squeezeup\squeezeup\squeezeup}. However, to make sure that the regulatory logic correctly reflects the desired topology, we use effective Boolean rules wherein no input is redundant or spurious \cite{PhysRevE.90.022815}. That is, for any input node $\mathit{i}$, $\mathit{f(\cdots,\sigma_{i}=1,\cdots)\neq f(\cdots,\sigma_{i}=0,\cdots)}$ for at least one pair of input configurations. We find by exhaustive enumeration that for effective rules, the probability $\mathit{\alpha(I_{j})}$ changes with the in-degree of the affected node $\mathit{j}$: $\mathit{\alpha(I_{j}}=1)=0, \mathit{\alpha(I_{j}}=2)=0.4,  \mathit{\alpha(I_{j}}=3) \approx 0.477, \mathit{\alpha(I_{j}}=4) \approx 0.498$. A Monte Carlo calculation shows that $\mathit{\alpha}$ approaches 0.5 as node in-degree increases. Thus one can readily see from the estimated average $\mathit{RP}$ (formula \ref{equ:RP}) that $\mathit{(\sum_{I_{j}}f(I_{j})\alpha(I_{j}))^{O_{i}}}$ decreases from 1 exponentially as $\mathit{O_{i}}$ increases from 0. Thus sink nodes and nodes with smaller out-degree have a greater contribution to the resilience probability, as they affect no or few other nodes. Given an average node out-degree, heterogeneity in the out-degree distribution tends to make the steady state of the network more stable against single node damage because it leads to more low-degree nodes. However, since $\mathit{\alpha(I_{j})}$ increases relatively slowly and saturates at 0.5 as $\mathit{I_{j}}$ increases, it is less straightforward to see the dependence between in-degree heterogeneity and the resilience probability of a steady state. We note that our mean-field approximation takes out-degree distribution and effective Boolean rules into consideration compared with annealed approximation.

We also analyze the effect of restricting the Boolean rules to nested canalizing rules, as research shows that the regulation in biological networks is frequently described in this way \cite{ref27}. A nested canalizing Boolean function with $\mathit{k}$ inputs can be generated by determining two sequences, the input sequence $\mathit{(I_{1},I_{2},\cdots,I_{k})}$ and the output sequence $\mathit{(O_{1},O_{2},\cdots,O_{k})}$, where $\mathit{I_{i}}$ or $\mathit{O_{i}}$ is either 0 or 1. The output $\mathit{o}$ as a function of input configuration $\mathit{(i_1,\cdots,i_k)}$ is thus determined through the hierarchy $\mathit{o=O_1}$ if $\mathit{i_1=I_1}$; $\mathit{o=O_2}$ if $\mathit{i_1\neq I_1}$ and $\mathit{i_2=I_2;}$ $\mathit{\cdots}$; $\mathit{o=O_k}$ if $\mathit{i_1\neq I_1}$, $\mathit{\cdots}$, $\mathit{i_{k-1}\neq I_{k-1}}$, $\mathit{i_k=I_k}$; $\mathit{ o=NOT\ O_k}$ if $\mathit{i_1\neq I_1}$, $\mathit{\cdots}$, $\mathit{i_{k-1}\neq I_{k-1}}$, $\mathit{i_k\neq I_k}$. The last condition is used to guarantee that the rule is an effective rule \cite{ref27}. All nested canalizing functions can be written in the above form up to a permutation of node order. We determine analytically, and verify by numerical simulations, that the probability that a node$'$s state will not change after knockout of one of its $\mathit{x}$ regulators is $\mathit{\alpha(x)=\frac{x-1}{x}}$ for nested Boolean functions generated by the method above with no bias in $\mathit{I_i}$  or $\mathit{O_i}$. This is because knocking out the first dominant canalizing variable $\mathit{i_1}$ (the probability of this is $\mathit{1/x}$), will change the input configuration; the output will be changed with probability 1/2, which is the probability that two outputs $\mathit{O_1}$ and $\mathit{O_l (l\neq1)}$ of the nested Boolean function hierarchy are different. Knocking out the second dominant canalizing variable changes the output only if $\mathit{i_1\neq I_1}$   (the probability of this is 1/2), the probability that output is changed is 1/2 as before under the condition $\mathit{i_1\neq I_1}$, and so on. Also notice that the order of the last two inputs in the hierarchy of the canalizing function does not affect the resilience probability, thus the probability of needing repair is $\frac{1}{x} (\frac{1}{2}+(\frac{1}{2})^2+\cdots+(\frac{1}{2})^{x-1}+(\frac{1}{2})^{x-1} )=\frac{1}{x}$, and therefore $\mathit{\alpha(x)}$ is $\mathit{\frac{x-1}{x}}$  . Notice that two different sequences may give the same rule, for example, for a one-input rule, $(\mathit{I_1}=1,\mathit{O_1}=1)$ is actually the same as $(\mathit{I_1}=0,\mathit{O_1}=0)$. Also, nested canalizing function ensembles generated by the input and output sequence with no bias lead to a different degeneracy of the Boolean functions in a Boolean table representation, in which the output of the Boolean function is specified for each possible input configuration. Simulations show that nested canalizing Boolean functions randomly picked from the Boolean table representation with equal probability have a different $\mathit{\alpha(x)}$ function: $\mathit{\alpha(x}=2)=0.5,\mathit{\alpha(x}=3)=0.625,\mathit{\alpha(x}=4)=0.712,\mathit{\alpha(x}=5)=0.766$. Regardless of the representation, $\mathit{\alpha}$ is larger for nested canalizing functions compared with random Boolean functions or effective random Boolean functions. This indicates that steady states of networks with nested Boolean functions will have an increased resilience against single node damage \cite{ref27,NC2,Peixoto2010}. Since $\mathit{\alpha}$ is smaller than 1, the conclusion that heterogeneity in the out-degree distribution tends to make the steady state of the network more resilient against single node damage holds for nested canalizing functions.

\squeezeup\squeezeup\squeezeup\squeezeup
\subsubsection{Damage probability in simulations of random network ensembles}
\label{sec:III.A.2}
\squeezeup\squeezeup\squeezeup
To estimate the resilience probability, we consider five random Boolean network ensembles with different in-degree/out-degree distributions, namely (a) constant in-degree and scale-free out-degree distribution (SF\textunderscore out), (b) constant in-degree and Poissonian out-degree (NK\textunderscore out), (c) constant in-degree and constant out-degree (NKK), (d) Poissonian in-degree distribution and constant out-degree (NK\textunderscore in), and (e) scale-free in-degree distribution and constant out-degree (SF\textunderscore in). The algorithm we used in generating these networks will give scale-free (power-law) degree distribution or Poisson degree distribution in the limit of very large network size. Even for small network sizes, the heterogeneity of these two types of networks is significantly different, e.g. the standard deviation of the first$'$s is approximately twice the second$'$s \footnote{When the average degree equals 2, the standard deviation of the node out-degree of one sample ensemble is 1.336 for Poisson distribution, 2.675 for scale-free distribution. When the average degree equals 3, the standard deviation is 1.588 for Poisson distribution, 3.188 for scale-free distribution.\squeezeup\squeezeup\squeezeup\squeezeup\squeezeup\squeezeup\squeezeup\squeezeup}.  For each ensemble, we generate 1000 networks with 20 nodes. To make sure that the generated ensemble has the desired topology and degree distribution, we only accept at least weakly connected networks and use effective rules when assigning a Boolean function to each node.  We study ensembles with average degree $\mathrm{<}K\mathrm{>}=$ 1, 2 and 3, which would be in frozen phase for $\mathrm{<}K\mathrm{>}=1$ and chaotic phase for $\mathrm{<}K\mathrm{>}=2$ or $3$ when considering the annealed approximation, the infinite network size limit and unbiased $\mathit{effective}$ Boolean rules \cite{ref30,ref29,ref28,semiannealed,PhysRevE55257}. Note that knowing the phase is not enough to predict the damage probability. For each network, we find all the steady states. For each steady state, we individually knock out (keep in the OFF state) every node that has the ON state in the steady state. A similar procedure can be followed to consider the constitutive expression (sustained ON state) of nodes that are currently OFF in the attractor; we do not explicitly consider this latter case. 

\begin{figure}[t]
\subfigure{
\includegraphics[width=3.5in]{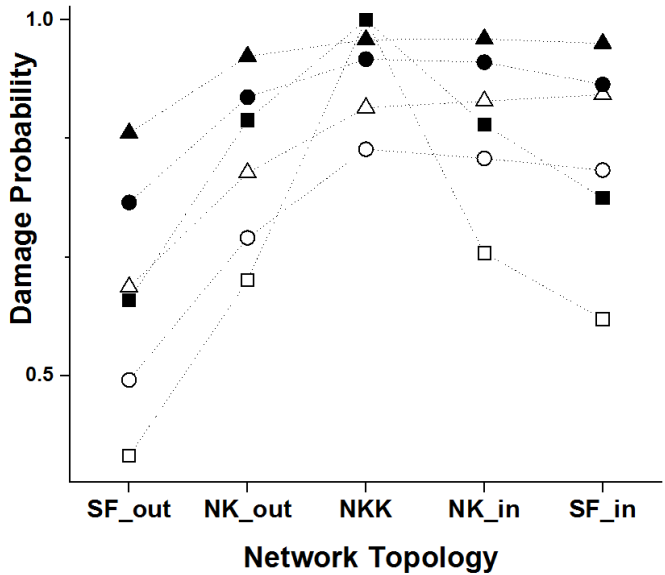}
}
\caption{The estimated damage probability across five ensembles with different degree heterogeneity. Different symbol shapes of the series represent different average degree,  $\mathrm{<}K\mathrm{>}=1$ (squares), $\mathrm{<}K\mathrm{>}=2$ (circles), and $\mathrm{<}K\mathrm{>}=3$ (triangles). Single node knockout results are shown with empty symbols and double node knockout results are shown with solid symbols. The standard error of the average damage probability is estimated to be in the order of 0.001, which is negligible compared with the size of the symbol.
}
\label{pic:fig1}
\squeezeup
\end{figure}

We estimate the resilience probability $\mathit{(RP)}$ and damage probability $\mathit{(DP=1-RP)}$ for networks with given topological characteristics by considering all steady states and all possible node knockouts with equal probability in the corresponding network ensemble. We do not weight our analysis based on the size of the attraction basin of steady states, as this property is not necessarily relevant in biological systems \cite{Albert20031}. Fig. \ref{pic:fig1} summarizes the simulation results for the estimated damage probability. In agreement with the theoretical result, for single node damage, given a fixed node in-degree, heterogeneity in out-degree leads to a smaller damage probability for the steady state (compare SF\textunderscore out, NK\textunderscore out and NKK results). In contrast, with node out-degree fixed, heterogeneity in in-degree distribution does not show a general trend and is connectedness dependent: the damage probabilities of the NKK, NK\textunderscore in and SF\textunderscore in ensembles are close for $\mathrm{<}K\mathrm{>}=2$ or $\mathrm{<}K\mathrm{>}=3$. Thus the theoretical analysis (see \ref{sec:III.A.1} $2^{nd}$ paragraph) is consistent with the computational result. A quantitative comparison of damage probability estimation by simulations and mean-field theory is shown in Table \ref{table:0} for selected ensembles. 

\squeezeup
\begin{table}[htbp]
\caption{A quantitative comparison of damage probability estimation }
\label{table:0}
\begin{tabular}{|l|l|l|l|l|l|}
\hline
Average degree/Method       & SF\textunderscore out & NK\textunderscore out & NKK & NK\textunderscore in & SF\textunderscore in \\ \hline
$\mathrm{<}K\mathrm{>}=2$\quad Simulation   & 0.494     & 0.694     & 0.818     & 0.805     & 0.789    \\ \hline
$\mathrm{<}K\mathrm{>}=2$\quad Mean-field    & 0.496     & 0.711     & 0.84     & 0.842     &  0.824    \\ \hline
$\mathrm{<}K\mathrm{>}=3$\quad Simulation    & 0.624    & 0.785     & 0.877     & 0.886     & 0.895    \\ \hline
$\mathrm{<}K\mathrm{>}=3$\quad Mean-field    & 0.609     & 0.805     & 0.891      & 0.905     &  0.906   \\ \hline
\end{tabular}
\caption*{ Single node damage probability estimation by simulations ($1^{st}$ and $3^{rd}$ row) and mean-field theory ($2^{nd}$ and $4^{th}$ row). Different columns correspond to different ensembles. Mean-field calculations employ the degree distributions of the generated ensemble.}
\squeezeup\squeezeup\squeezeup\squeezeup
\end{table}

\squeezeup\squeezeup\squeezeup\squeezeup\squeezeup
\subsection{Double node damage}
\squeezeup\squeezeup
\subsubsection{Classification of the resilience scenarios of double node damage}
\label{sec:III.B.1}
\squeezeup\squeezeup\squeezeup
In this section, we generalize the single node damage repair algorithm proposed in Ref. \cite{ref1} and investigate the properties of interventions that prevent the cascading effect of knocking out two nodes in a network. The potential combinatorial effects of simultaneous damage to two nodes have been named genetic interactions in biological systems \cite{ref10}. A specific example of cases where combined knockout of two genes has a stronger effect than the sum of the effects of the individual knockouts is synthetic lethality \cite{syntheticlethality1,syntheticlethality2,syntheticlethality3}. The converse case was termed synthetic viability  \cite{syntheticlethality3,syntheticlethality4}. Both of these genetic interactions have been studied experimentally \cite{syntheticlethality1,syntheticlethality2} and theoretically \cite{syntheticlethality3,syntheticlethality4}. Here we go beyond the identification of genetic interactions by determining the specific edge additions through which the cascading effects of cumulative damage can be prevented.


When repair is necessary, for each sensitive node, we define candidate nodes as nodes that are neither its pre-existing regulators nor the sensitive node itself, and we add a suitable interaction starting from a candidate node to prevent the state change. (We avoid using pre-existing regulators since it is less biologically feasible \cite{ref1}.) This way, we preserve the steady state aside from the immediate impact on the damaged node and block the cascading effect as soon as possible. Specifically, say node $\mathit{i}$ is regulated by nodes that belong to set $\mathit{A}$, $\mathit{x_i=f(x_{j_1 },\cdots,x_{j_k } )}$, where $\mathit{j_1,\cdots,j_k\in A}$. If one wants to repair node $\mathit{i}$ so that it remains ON ($\mathit{x_i}=1$), one needs to find a candidate node $\mathit{l}$ (i.e. $\mathit{l\notin A}$ and $\mathit{l\neq i}$) and modify the rule such that $\mathit{x_i=f(x_{j_1 },\cdots,x_{j_k } )\ OR\ x_l}$  if $\mathit{x_l=}1$ or such that $\mathit{x_i=f(x_{j_1 },\cdots,x_{j_k } )\ OR\ ( NOT\ x_l )}$  if $\mathit{ x_l=}$0. Here $\mathit{AND}$, $\mathit{OR}$ and $\mathit{NOT}$ are Boolean functions. Similarly, if one wants to repair node $\mathit{i}$ so that it remains OFF $\mathit{(x_i=0)}$, one can modify the rule to be  $\mathit{x_i=f(x_{j_1 },\cdots,x_{j_k } )\ AND\ ( NOT\ x_l )}$ if $\mathit{x_l}=1$; or  $\mathit{x_i=f(x_{j_1 },\cdots,x_{j_k } )\ AND\ x_l}$ if $\mathit{x_l}$=0 \cite{ref1}. Assuming that a candidate node with the  appropriate $\mathit{x_l}$  value exists, which is generally the case in realistic networks, regulation of this sort is always possible in principle \cite{ref1}.  We say that  a repair solution exists if each sensitive node can be repaired. In the algorithm for double node damage, the sensitive node set is determined after knockout of both nodes;  then for each sensitive node, a candidate node set is identified with the additional restriction of excluding both damaged nodes. Then, similarly to single node knockout, an interaction is added from an appropriate candidate node to each sensitive node. 
\begin{table}[!tbp]
\squeezeup\squeezeup
\caption{Classifications of different situations comparing single node damage and double node damage.}
  \label{table:1}
\begin{tabular}{|l|l|l|}
\hline
\begin{tabular}[c]{@{}l@{}}Class/\\ subclass\end{tabular} & \begin{tabular}[c]{@{}l@{}}Status of SS after\\  single node damage\end{tabular}  & \begin{tabular}[c]{@{}l@{}}Status of  SS after \\ double node damage\end{tabular} \\ \hline
1 (b)                                                     & stable for both                                                                         & stable                                                                                      \\ \hline
2 (c)                                                     & stable for both                                                                          & Needs repair                                                                            \\ \hline
3 (a)                                                     & \begin{tabular}[c]{@{}l@{}}stable in one case,\\ needs repair in the other case\end{tabular} & stable                                                      \\ \hline
\begin{tabular}[c]{@{}l@{}}4 \\ (a,b,c,d)\end{tabular}         & \begin{tabular}[c]{@{}l@{}}stable in one case,\\ needs repair in the other case\end{tabular} & Needs repair            \\ \hline
5 (a)                                                     & Both need repair                                                                         & stable                                                                                      \\ \hline
\begin{tabular}[c]{@{}l@{}}6 \\ (a,b,c,d)\end{tabular}         & Both need repair                                           & Needs repair                                                                            \\ \hline
a                                                         & \multicolumn{2}{l|}{$\mathit{S_A \cup S_B  \supsetneq S_{AB}}$}                                                \\ \hline
b                                                         & \multicolumn{2}{l|}{$\mathit{S_A \cup S_B  = S_{AB}}$}                                                             \\ \hline
c                                                         & \multicolumn{2}{l|}{$\mathit{S_A \cup S_B  \subsetneq S_{AB}}$}                                                 \\ \hline
d                                                         & \multicolumn{2}{l|}{$\mathit{(S_A \cup S_B) \backslash  S_{AB} \neq \emptyset \quad and\quad S_{AB} \backslash (S_A \cup S_B)  \neq \emptyset} $}                          \\ \hline
\end{tabular}
\caption*{The first column is the class and its possible subclasses, whose definitions are given in the last four rows. The subclasses are defined based on the relationships between $\mathit{S_{AB}}$ and $\mathit{S_A \cup S_B}$. $\mathit{S_A \cup S_B  \supsetneq S_{AB}}$   means $\mathit{S_{AB}}$ is a true subset of $\mathit{S_A \cup S_B}$. $\mathit{(S_A \cup S_B) \backslash  S_{AB} \neq \emptyset}$ and $\mathit{S_{AB} \backslash (S_A \cup S_B)  \neq \emptyset} $  means that $\mathit{S_{AB}}$  is not a subset of $\mathit{S_A \cup S_B}$ and $\mathit{S_A \cup S_B}$ is also not a subset of $\mathit{S_{AB}}$, where \textbackslash \enspace means relative complement. SS means steady state.
}
\squeezeup\squeezeup\squeezeup\squeezeup
\end{table}

When considering the damage of node A, damage of a different node B, and damage of both nodes, six outcomes are possible, which are summarized in the first six rows of Table \ref{table:1}. These six outcomes have been discussed in the context of synthetic lethality in random threshold networks \cite{syntheticlethality3}. In order to compare the repair solutions, we denote the sensitive node sets after damage to node A, B, and both A and B as $\mathit{S_A, S_B,}$ and $\mathit{S_{AB}}$, respectively. If no node is a child node of both node A and node B, $\mathit{S_{AB}=S_A \cup S_B}$. However, if a node is a child node of both nodes A and B, different situations can emerge as indicated in the last four rows of Table \ref{table:1}. For completeness, for each class (i.e., situation) we list the possible subclasses in the first column of the table. Classes 1, 2, 3, and 5 admit a single subclass only, while classes 4 and 6 can have any of the four subclasses.

\begin{table}[tbp]
\squeezeup\squeezeup
\caption{The ten two-input effective Boolean functions and their classification in double node damage}
\label{table:2}
\begin{tabular}{|l|l|l|l|l|l|}
\hline
Function Name       & (1,1) & (1,0) & (0,1) & (0,0) & Class \\ \hline
(NOT A) OR (NOT B)  & 0     & 1     & 1     & 1     & 6b    \\ \hline
(NOT A) OR B        & 1     & 0     & 1     & 1     & 3a    \\ \hline
A OR (NOT B)        & 1     & 1     & 0     & 1     & 3a    \\ \hline
A OR B              & 1     & 1     & 1     & 0     & 2c    \\ \hline
XNOR(A,B)           & 1     & 0     & 0     & 1     & 5a    \\ \hline
XOR(A,B)            & 0     & 1     & 1     & 0     & 5a    \\ \hline
A AND B             & 1     & 0     & 0     & 0     & 6b    \\ \hline
A AND (NOT B)       & 0     & 1     & 0     & 0     & 3a    \\ \hline
(NOT A) AND B       & 0     & 0     & 1     & 0     & 3a    \\ \hline
(NOT A) AND (NOT B) & 0     & 0     & 0     & 1     & 2c    \\ \hline
\end{tabular}
\caption*{The first column indicates the name of the function of A and B, where XOR(A,B)= (A AND (NOT B)) OR ((NOT A) AND B), XNOR(A,B)=NOT XOR(A,B). The second to fifth columns list the value of the respective function for input combinations (A=1, B=1), (A=1, B=0), (A=0, B=1), (A=0, B=0). The sixth column gives the classification (as in Table \ref{table:1}) of a network motif composed of three nodes, wherein A and B are source nodes and the regulation of C is described by the respective Boolean function of A and B. This classification assumes that the input nodes are $A=1, B=1$ initially. The letter in the last column gives the subclass based on the relationship between $\mathit{S_A \cup S_B}$  and $\mathit{S_{AB}}$ as described in Table \ref{table:1}. }
\squeezeup\squeezeup\squeezeup\squeezeup\squeezeup
\end{table}

\begin{figure*}[htbp]
\squeezeup\squeezeup\squeezeup\squeezeup
\subfigure{
\includegraphics[width=6.5in]{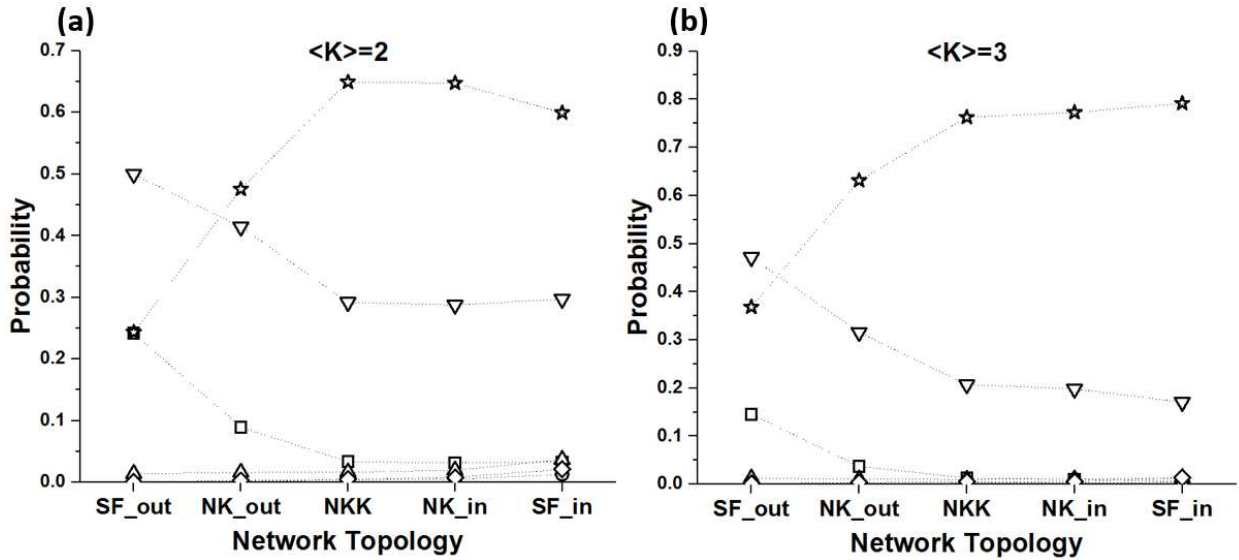}
}
\caption{Probability of each class of double node knockout across the five ensembles with different degree distributions. The left and right graph shows the result for networks with average degree $\mathrm{<}K\mathrm{>}=2$ and $\mathrm{<}K\mathrm{>}=3$ respectively. Classes 1 to 6 are drawn in square, circle, up triangle, down triangle, diamond, and star symbols respectively.
}
\label{pic:fig2}
\squeezeup
\end{figure*}

In order to gain insight into how different networks lead to the different outcomes of Table \ref{table:1}, we consider a simple network motif, in which two nodes (A, B) regulate a third node C. We determine which class and subclass each two-variable Boolean function belongs (Table \ref{table:2}). We start from state (1, 1) for the two nodes (in the order A, B). If the output of state (0,1) is different from that of state (1,1), the state needs repair after damage to node A; similar conclusions apply to all the cases. The symmetrical AND rule and its negation belong to class 6, the symmetrical OR rule and its negation belong to class 2, four cases of unsymmetrical two-variable regulation belong to class 3, and the XOR/XNOR functions belong to class 5. The three node motif holds the same properties when embedded within a larger network. However, we emphasize   that a network containing a three-node motif and additional nodes does not necessarily fall into the same category as the three-node motif alone, since different parts of the network may be different for different damage situations. 

We find that in subclass b, that is, $\mathit{S_A \cup S_B  = S_{AB}}$, the repair solution for the double node damage will always be a subset of the ``direct product'' of the single node damage repair solution. More rigorously, let $\mathit{S_A=\{A_1,\cdots,A_m\}}$ and $\mathit{S_B=\{B_1,\cdots,B_n \}}$. A repair solution after knocking out node A has the form $\mathit{(r_{A_1},\cdots,r_{A_m} )}$, where $\mathit{r_{A_1}}$ represents a way to stabilize node $\mathit{A_1}$. Let $\mathit{R_{A_1 }}$ be the set containing all $\mathit{r_{A_1}}$ that appears in all possible repair solutions. The set of all possible solutions after knocking out node A will be denoted as $\mathit{G_A=\{(r_{A_1 },\cdots,r_{A_m } ):r_{A_i }\in R_{A_i} \}}$. For the direct product of single node damage repair solution, $\mathit{S_D=\{D_1,\cdots,D_p\}}$, $\mathit{p=| S_A \cup S_B |}$; $\mathit{R_{D_i }=R_{A_i }\cup R_{B_i }}$ if $\mathit{D_i\in S_A\cap S_B}$ and  $\mathit{D_i=A_i=B_i}$;  $\mathit{R_{D_i }=R_{A_i }}$ if $\mathit{D_i\in S_A\cap S_B^c}$ and $\mathit{D_i=A_i}$; $\mathit{R_{D_i }=R_{B_i }}$ if $\mathit{D_i\in S_A^c \cup S_B}$  and $\mathit{D_i=B_i}$. Then the direct product of single node damage repair solution is given by $\mathit{G_D=\{(r_{D_1 },\cdots,r_{D_m } ):r_{D_i }\in R_{D_i }\}}$.  This can be explained in the following way: since a node only has two states in a Boolean network, it will either be stable or will need repair. When the node needs repairing, damage to an additional node reduces the number of candidate nodes that can be used as starting points of the repair edges; nothing else should happen. Thus, the individual single node repair solutions are compatible with each other. Another observation is that $\mathit{S_{AB}=S_A=S_B}$ can only happen if A and B are regulating the same node(s). Otherwise, part of the damage, and thus also of the repair solutions, would be independent of each other.

\squeezeup\squeezeup\squeezeup\squeezeup
\subsubsection{Damage probability and class distribution in simulations of random network ensembles}
\squeezeup\squeezeup\squeezeup
Similarly to Sec. \ref{sec:III.A.2}, we study the effect of degree heterogeneity on the resilience probability in a double knockout setting. We also explore the distribution of the repair categories introduced in Sec. \ref{sec:III.B.1} using simulations of random Boolean networks. The computational details are similar as in Sec. \ref{sec:III.A.2} except we consider all possible pairs of knockouts to obtain the estimation for the damage probability and the classification of each category.

As shown in Fig. \ref{pic:fig1}, the damage probability after double knockout is rather high regardless of the degree distribution and is higher than the damage probability after single knockout. As one can see, the double-knockout damage probability is higher in a network with higher average degree, which is consistent with the established conclusion that the complexity of the dynamics increases with larger average node in-degree \cite{ref28,ref29,ref30}. The NKK model with K=1 is an exception; here the damage probability is 1 whether one or two nodes are damaged. This is because this network forms a single cycle. The only possible effective Boolean rules for K=1 are the identity (the output equals the input) and the negation. Thus knocking out any currently-ON node in the network will induce a change in its child node, which means the network will need to be repaired. 

\begin{figure*}[htbp]
\squeezeup\squeezeup\squeezeup\squeezeup
\subfigure{
\includegraphics[width=6.5in]{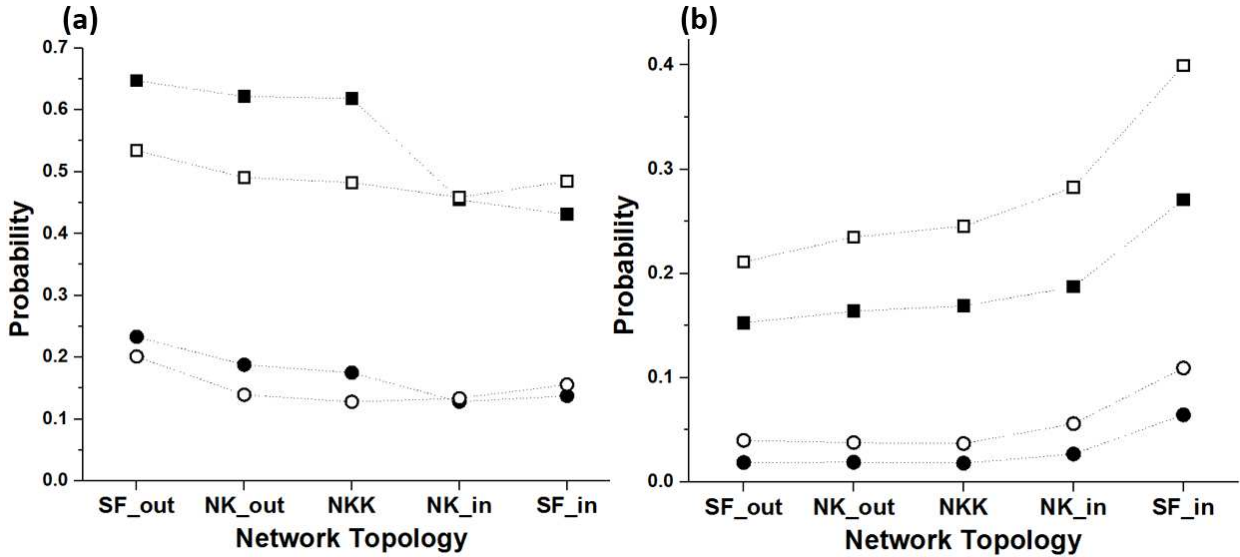}
}
\caption{The probability that one needs to repair more or different (circles) or fewer (squares) nodes for simultaneous damage of two nodes compared to the union of the repairs needed for individual damage to each of the nodes. The probability that one needs to repair the same nodes is 1 minus the sum of the two shown probabilities. The same network ensembles as in Fig. \ref{pic:fig1} and \ref{pic:fig2} are used. Solid symbols represent $\mathrm{<}K\mathrm{>}=2$ results and empty symbols represent $\mathrm{<}K\mathrm{>}=3$. (a)  Simultaneous damage of two nodes that share a downstream target. (b) The general case of simultaneous damage of two nodes.
}
\label{pic:fig3}
\squeezeup
\end{figure*}

Based on the simulations, the damage probabilities of ensembles with fixed out-degree (K=2 and K=3) for double node knockout are rather close to each other; in-degree heterogeneity does not significantly change the damage probability. However, when we compare the three ensembles with fixed in-degree, out-degree heterogeneity leads to a decrease in the damage probability; this is because of the abundance of sink nodes. These results are similar to the results of the single node knockout.

To illustrate the distribution of the double damage classes introduced in Table \ref{table:1}, in Fig. \ref{pic:fig2} we plot the probability of each class in the five ensembles. Based on the simulations, class 2 (both single damage cases are stable, repair is needed for double damage), class 3 (repair is needed for one case of single damage, stable after double damage) and class 5 (repair is needed for each single damage, stable after double damage) have very low probability of occurrence. The reason is that the occurrence of these situations requires that the nodes being knocked out are regulating a common target. In contrast, most randomly chosen node pairs are independent. Class 6 (see Table \ref{table:1})  tends to have the highest probability, followed by class 4  and class 1 ; the probability of these cases also varies more in the different ensembles. Comparing the three ensembles with a fixed out-degree (K=2 or K=3), the probability of each class is fairly close according to the simulation. Comparing the three ensembles with a fixed node in-degree, we can readily see that heterogeneity in node out-degree leads to a smaller probability for class 6 (stars) and larger probability for classes 4 (down triangles) and 1 (squares). This is related to the fact that heterogeneity in node out-degree leads to more sink nodes in the network.     

\begin{figure*}[htbp]
\squeezeup\squeezeup\squeezeup\squeezeup
\subfigure{
\includegraphics[width=6.5in]{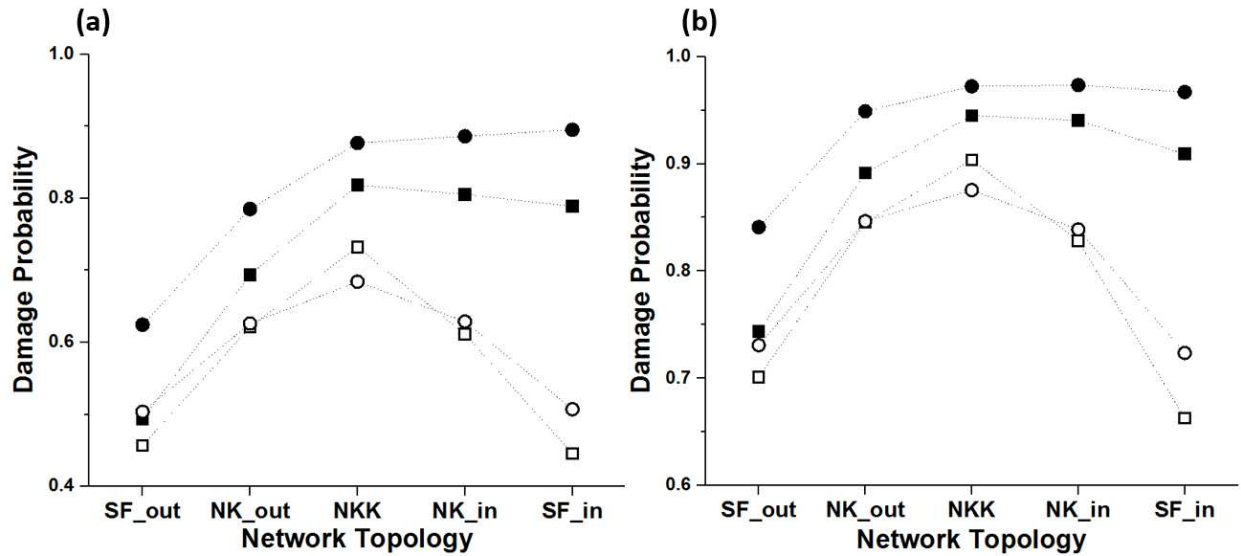}
}
\caption{Damage probability in network ensembles using effective Boolean rules (solid symbols) or nested canalizing rules (empty symbols). Square symbols represent $\mathrm{<}K\mathrm{>}=2$ and circular symbols represent $\mathrm{<}K\mathrm{>}=3$. (a) Single node knockout; (b) double node knockout.
}
\label{pic:fig4}
\end{figure*}

As we are interested in combinatorial effects of double node knockout, we marginalize all the (sub)classes into three categories based on whether we need to repair more or different nodes (class 2, 4c, 4d, 6c, and 6d), the exact same set of nodes (class 1, 4b and 6b), or less nodes (classes 3, 5, 4a, and 6a) in case of double damage compared to the union of the two single damage cases. We estimate the probability of each category by simulation using the five ensembles. If the two nodes being knocked out do not share a target, the two damage processes are independent and there will not be any combinatorial effect. It is therefore of particular interest to calculate the probability of each category in just the cases wherein the two nodes share a target (Fig. \ref{pic:fig3}(a)), and compare with the general case (Fig. \ref{pic:fig3}(b)). Since prior research shows that the average degree of biological networks is around 2 \cite{Kauffman1,Kauffman2}, we focus on $\mathrm{<}K\mathrm{>}=2$ and $\mathrm{<}K\mathrm{>}=3$.
According to both Fig. \ref{pic:fig3}(a) and \ref{pic:fig3}(b), the probability that we need to repair fewer nodes (subclass a) in double knockouts is larger than the probability that we need to repair more or different nodes (subclasses c and d). This is consistent with the fact that in Table \ref{table:2}, there are more motifs corresponding to subclass a than to subclass c. For $\mathrm{<}K\mathrm{>}=2$, compared with networks with constant in-degree (the left three ensembles in Fig.  \ref{pic:fig3}(a)), networks with constant out-degree and heterogeneous in-degree distribution (the right two ensembles in Fig.  \ref{pic:fig3}(a)) demonstrate a lower probability for cases wherein one needs to repair more or fewer nodes; for $\mathrm{<}K\mathrm{>}=3$, the probabilities are close to each other. As the network topology changes from constant in-degree and scale-free out-degree distribution to constant in-degree and out-degree to constant out-degree and scale-free in-degree distribution (from left to right in Fig.  \ref{pic:fig3}(b)), the percentage of node pairs sharing a target node among all possible pairs increases. This change is more dramatic than the change in the probability of the three categories across different ensembles in Fig.  \ref{pic:fig3}(a). Thus as shown in Fig.  \ref{pic:fig3}(b), the probability of cases wherein one needs to repair more nodes (circles) or less nodes (squares) among all node pairs increases across the five ensembles.

As discussed in Sec. \ref{sec:III.A.1}, using nested canalizing functions helps make a steady state more resilient to single node damage under the same network topology. This is confirmed by simulation results summarized in Fig. \ref{pic:fig4}(a). The damage probability is smaller for networks with nested canalizing functions (empty symbols) for all five ensembles with $\mathrm{<}K\mathrm{>}=2$ or $\mathrm{<}K\mathrm{>}=3$. This conclusion holds for double node damage, as shown in Figure  \ref{pic:fig4}(b). As discussed in Sec.  \ref{sec:III.A.1}, out-degree heterogeneity leads to lower damage probability for both effective and canalizing functions (compare the first three ensembles). In the case of nested canalizing Boolean functions, in-degree heterogeneity also leads to a lower damage probability, in contrast with its minor effect in case of effective Boolean functions. This is because higher in-degree leads to more stability for nested canalizing functions, reflected in the fact that the stability probability $\mathit{\alpha(I_j)}$ of nested canalizing functions keeps increasing steadily and is much larger than that of effective Boolean functions for higher $\mathit{I_j}$ (see  Sec. \ref{sec:III.A.1}).

\squeezeup\squeezeup\squeezeup\squeezeup
\subsection{Single node damage in networks with two steady states}
\squeezeup\squeezeup
\subsubsection{General Discussion}
\squeezeup\squeezeup\squeezeup

\begin{figure*}[!tbp]
\squeezeup\squeezeup\squeezeup\squeezeup
\subfigure{
\includegraphics[width=6.5in]{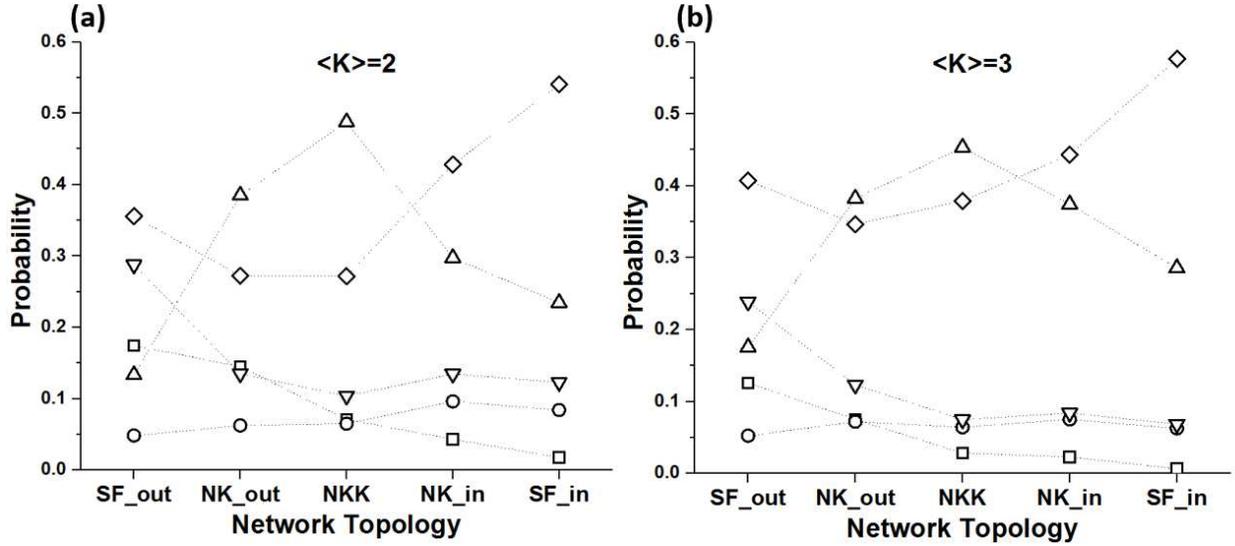}
}
\squeezeup\squeezeup
\caption{Probability of outcomes of node knockout on pairs of steady states across five network ensembles with different degree distributions. The left and right graph shows the result for networks with average degree $\mathrm{<}K\mathrm{>}=2$ and $\mathrm{<}K\mathrm{>}=3$ respectively. The classes are grouped as 1 (squares),  the union of 2 and 3 (circles), the union  of 4, 5, 6 (up triangles),  7 (down triangle), and union of 8 and 9 (diamonds). 
}
\label{pic:fig7}
\squeezeup
\end{figure*}

Another follow-up direction is to explore the effect of single node damage on two different steady states of a network. The goal is to see whether a single solution can remedy the damage in multiple attractors (steady states here) at the same time. To classify all the situations of knockout damage to a single node, we observe that the damaged node may be normally (when undamaged) ON in both steady states, or ON in one steady state and OFF in the other. (We do not consider the situation that the node is OFF in both steady states, as the knockout damage will not change anything to either steady state). The categorization of the constitutive expression type damage will be analogous.

If the node is ON in both steady states, the steady states can be both stable, both in need of repair or one is stable and the other needs repair. If the node has different states in the two steady states, only one of them needs repair, as summarized in Table \ref{table:3}.

We explore the probability distribution of the classification shown in Table \ref{table:3} in random Boolean networks. The computational details are similar to Sec. \ref{sec:III.A.2} except we consider all possible single node knockouts for every pair of steady states for a specific network. As there are 9 classes and each class may have a small probability, we marginalize class 2 and 3 (where the node is ON in both steady states before damage and one of the steady states needs repair after damage), class 4, 5 and 6 (where both need repair), class 8 and 9 (where node has different states before damage and one of the steady states needs repair after damage). As shown in Fig. \ref{pic:fig7}, we found that the class in which both steady states need repair (up triangle) is less probable in heterogeneous networks. The class in which both steady states are stable  (squares) is more probable in out-degree heterogeneous networks as sink nodes contribute to the resilience probability of steady state as in Sec. \ref{sec:III.A}.

\begin{table}[htbp]
\centering
\caption{Classification of the outcomes of single node knockout for two steady states of a network}
\label{table:3}
\begin{tabular}{|l|l|l|}
\hline
\begin{tabular}[c]{@{}l@{}}Situation\\  Index\end{tabular} & \begin{tabular}[c]{@{}l@{}}Node state \\ before damage\\ in the two SSs\end{tabular} & \begin{tabular}[c]{@{}l@{}}Status of the two SSs \\ after node damage\end{tabular}                                  \\ \hline
1                                                          & ON in both                                                                           & Both stable                                                                                                         \\ \hline
2                                                          & ON in both                                                                           & \begin{tabular}[c]{@{}l@{}}One SS is stable,\\ the other SS needs repair,\\ compatible solutions exist\end{tabular} \\ \hline
3                                                          & ON in both                                                                           & \begin{tabular}[c]{@{}l@{}}One SS is stable, \\ the other SS needs repair, \\ no compatible solutions\end{tabular}  \\ \hline
4                                                          & ON in both                                                                           & \begin{tabular}[c]{@{}l@{}}Both SSs need repair, \\ common solution(s) exist\end{tabular}                           \\ \hline
5                                                          & ON in both                                                                           & \begin{tabular}[c]{@{}l@{}}Both need repair, \\ compatible (but not common) \\ solution(s) exist\end{tabular}       \\ \hline
6                                                          & ON in both                                                                           & \begin{tabular}[c]{@{}l@{}}Both need repair, \\  no compatible solutions\end{tabular}                               \\ \hline
7                                                          & ON, OFF                                                                              & Both stable                                                                                                         \\ \hline
8                                                          & ON, OFF                                                                              & \begin{tabular}[c]{@{}l@{}}SS with node ON needs repair,\\ compatible solutions exist\end{tabular}                  \\ \hline
9                                                          & ON, OFF                                                                              & \begin{tabular}[c]{@{}l@{}}SS with node ON needs repair,\\ no compatible solutions\end{tabular}                     \\ \hline

\end{tabular}
\caption*{The table lists nine situations that will happen after knocking out a single node in the network that has two steady states (SSs). The second column indicates the state of the knocked-out node before damage. The third column specifies whether the damaged steady state is stable against the damage or needs repair. We use the term ``common solution'' for when we need to repair exactly the same set of nodes for the two steady states after single node knockout, we say ``compatible solution'' for when there exists a solution that can be used to stabilize both steady states. Thus all common solutions are compatible solutions.}
\squeezeup\squeezeup\squeezeup\squeezeup\squeezeup\squeezeup\squeezeup\squeezeup\squeezeup
\end{table}

We are particularly interested in determining whether or not there are common repair solutions in the cases where both steady states need repair (classes 4, 5 and 6). We find from our simulations on network ensembles that not having common solutions is less probable (the fraction of classes 5 and 6 is between 5$\%$ and 7$\%$ for different ensembles with $\mathrm{<}K\mathrm{>}=2$), thus we enumerate these situations. Similar to Sec. \ref{sec:III.B.1}, we start by looking for three-node motifs that will lead to no common solutions. Reexamining the 10 motifs in Table \ref{table:2}, and considering pairs of possible steady states for these motifs, we find that only the XOR/XNOR motif will forbid a common solution for repairing the two steady states.  The XOR/XNOR motif is rarely observed in biological networks, as they represent cases where each regulator can switch between being an activator or inhibitor depending on the state of the other regulator.

\begin{figure*}[!t]
\squeezeup\squeezeup\squeezeup\squeezeup
\subfigure{
\includegraphics[width=6.5in]{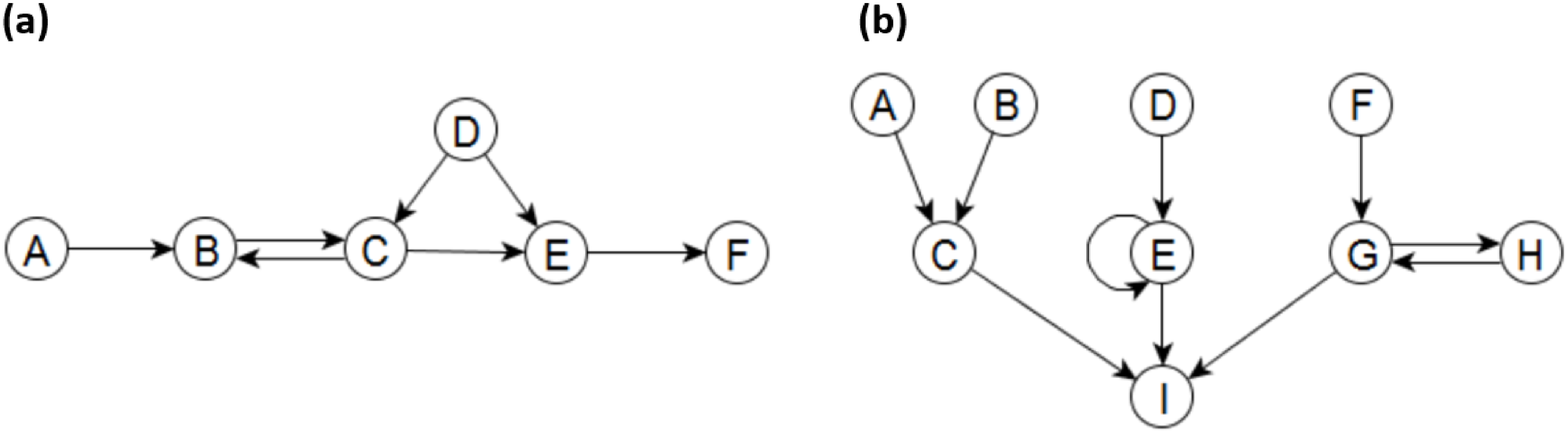}
}
\squeezeup\squeezeup
\caption{(a) A simple network to illustrate the situation when there are no valid repair candidates. All the edges are positive, the updating rules are B = A AND C, C = B OR D, E = C AND D. There are two steady states, (1, 1, 1, 1, 1, 1) and (1, 1, 1, 0, 0, 0). If node A is knocked out, we need to repair node B.  All candidate nodes (namely D, E and F) are in different states in the two steady states, thus no common solutions exist. (b) An example network to illustrate incompatibility in stabilizing two steady states at the same time because the two steady states only differ in the state of the knocked-out node and of the sensitive node. C = A OR B, E = D OR E, G = F OR H, I = A OR E OR G. First, we consider the effect of knockout of source node A on the steady state pair (1, 0, 1, 1, 1, 1, 1, 1, 1) and (0, 0, 0, 1, 1, 1, 1, 1, 1).  When knocking out node A, we need to repair node C to be ON for the first steady state.  However, fixing C to be ON  will eliminate the other steady state (where C is OFF) as all the candidate nodes (B, D, E, F, G, H or I) have  the same state in the two steady states; thus no compatible repair solutions exist. The incompatibility mechanism is the same for knockout of node E in case of the steady state pair (0, 0, 0, 0, 1, 0, 0, 0, 1) and (0, 0, 0, 0, 0, 0, 0, 0, 0), and the knockout of node G in case of the state pair (1, 1, 1, 1, 1, 0, 1, 1, 1) and (1, 1, 1, 1, 1, 0, 0, 0, 1). 
}
\label{pic:fig8}
\squeezeup
\end{figure*}

Another mechanism that will lead to no common solutions for repairing two steady states is that there is no valid candidate to use as a starting point of an additional edge. One such situation is that all the candidate nodes have different states in the two steady states, thus none of them can be used to realize the same function in the two steady states. This is exemplified in Fig. \ref{pic:fig8}(a).  Another situation is that the  sensitive node  is regulated by almost every node (other than the node itself) and there are no nodes left to be repair candidates since current regulators cannot be used. A combination of the two situations can also lead to no valid candidate for repair.

When the node is ON in one steady state and OFF in the other steady state, the damage will do nothing to the second steady state. However, the repair solution for the first steady state may or may not be compatible with the second steady state. Our simulations using random networks suggest that the incompatible situation is rarer. Incompatibility can arise in a lot of simple motifs of two or three nodes, including a single regulating edge (positive or negative), OR gate, AND gate, XOR gate, XNOR gate. The reason why this situation is rare in a real network is that if the network has nodes that have different states in the two steady states, any of these nodes can be used as starting points to an additional edge to node B. This additional edge will have an opposite effect in the two steady states and thus it can solve the incompatibility problem.
It is rare, but still possible, that two steady states of a network have the same state for most of nodes and only differ in the state of the knocked-out node, the sensitive node, and possibly its current regulators. This can happen if the knocked-out node is part of a bistable motif connected with the rest of the network with a canalizing function such as an OR gate. Thus the bistable motif neither affects nor is affected by the rest of the network in a steady state. Examples of bistable motifs are a source node, a node with a self-loop and a two-node feedback loop (see Fig. \ref{pic:fig8}(b)).

\squeezeup\squeezeup\squeezeup\squeezeup
\subsection{Biological Examples}
\squeezeup\squeezeup\squeezeup
In this section, we apply our algorithms to study the resilience of the T cell large granular lymphocyte (T-LGL) leukemia network \cite{ref31} and the epithelial-to-mesenchymal transition (EMT) network \cite{ref33}. 

\begin{figure}[!tbp]
\squeezeup\squeezeup\squeezeup\squeezeup
\subfigure{
\includegraphics[width=2.5in,left]{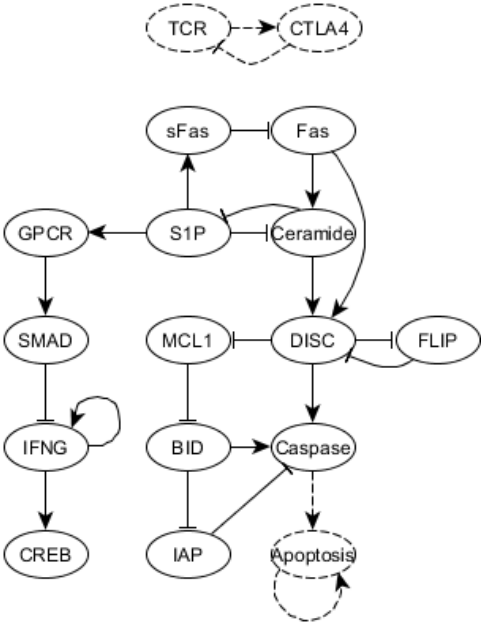}
}
\caption{Reduced T-LGL leukemia signaling network. An arrowhead or flat bar end indicates a positive or negative regulation edge respectively. The nodes and edges drawn in dashed lines are ignored in the analysis (see Appendix \ref{appendix:0}). 
}
\label{pic:fig5}
\squeezeup\squeezeup
\end{figure}

T-LGL leukemia is a rare blood cancer. While normal T cells undergo activation induced cell death (apoptosis) after successfully fighting a virus, leukemic T-LGL cells survive. The network model  constructed by Zhang \textit{et al.} \cite{ref31} includes the proteins involved in the activation of T cells, in activation induced cell death, as well as a number of proteins that were observed to be abnormally highly expressed or active in T-LGL cells. The model describes the regulation of each of these proteins with Boolean rules, and captures the normal (apoptosis) and leukemic (survival) states of the system \cite{ref31}. The original network has 60 nodes, including three source nodes, and 142 regulatory edges. By fixing all the states of source (unregulated) nodes in the biologically relevant condition and iteratively replacing fixed node states in the Boolean rules, one can reduce the network to a smaller network, whose nodes$'$ states are not determined by the source nodes alone but rather by the specific dynamic trajectory of the system \cite{ref32}. We perform additional network simplification as specified in Appendix \ref{appendix:0}. The reduced network model (Fig. \ref{pic:fig5}) has two steady states, namely a disease (T-LGL) state (0, 0, 0, 0, 0, 1, 0, 1, 1, 0, 1, 1, 1, 1) and a normal T cell state committed to the path to apoptosis (1, 0, 1, 1, 1, 0, 1, 0, 0, 0, 0, 0, 0, 0), where the nodes are in the alphabetic order, BID, CREB, Caspase, Ceramide, DISC, FLIP, Fas, GPCR, IAP, IFNG, MCL1, S1P, SMAD, sFas.

\begin{figure}[!tbp]
\squeezeup\squeezeup\squeezeup\squeezeup
\subfigure{
\includegraphics[width=4.5in,right]{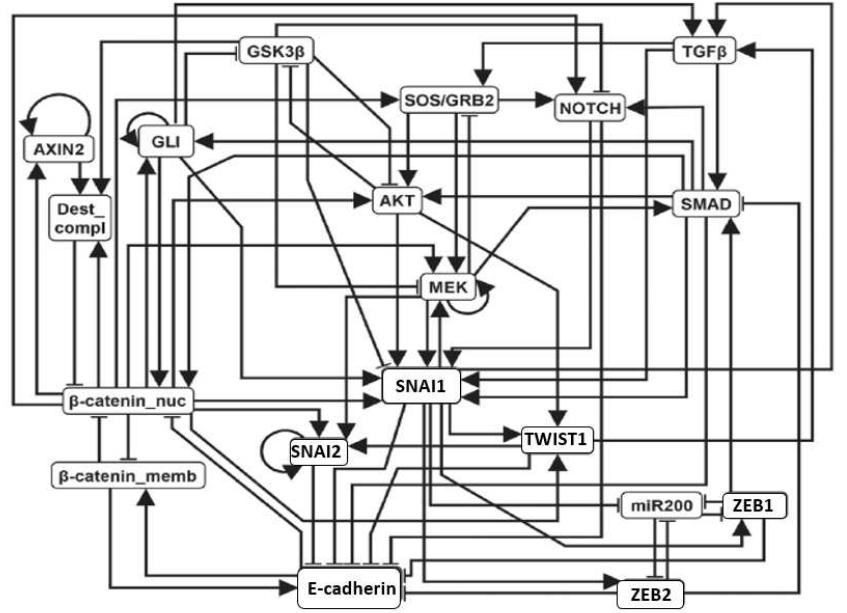}
}
\caption{Reduced EMT Network. Nodes represent proteins and miRNAs involved. An arrowhead or flat bar end indicates positive or negative regulation, respectively. 
}
\label{pic:fig6}
\squeezeup\squeezeup
\end{figure}

EMT is a cell fate change involved in embryonic development which can be reactivated during cancer metastasis \cite{ref33}. During EMT, epithelial cells lose their original adhesive property, leave their primary site, invade neighboring tissue, and migrate to distant sites as mesenchymal cells. A Boolean network model of EMT in the context of hepatocellular carcinoma invasion has been established by Steinway \textit{et al.} \cite{ref34}.  The EMT network has 70 nodes and 135 edges. Steinway \textit{et al.} performed a network reduction to obtain a network with 19 nodes and 70 edges (Fig. \ref{pic:fig6}). This type of network reduction has been shown to have no effect on the permitted dynamics and enables us to fully explore the state space \cite{ref34}. In the reduced network, the adhesion factor E-cadherin is the sink node and its OFF state will indicate the transition to a mesenchymal state. The reduced network has two steady states, the epithelial state (0, 1, 1, 0, 0, 1, 0, 1, 0, 0, 0, 0, 1, 0, 0, 0, 0, 0, 1) and the mesenchymal state (1, 1, 0, 1, 0, 0, 1, 0, 1, 1, 1, 1, 1, 0, 1, 1, 1, 1, 0), written in the order of AKT, AXIN2, $\beta$-catenin\textunderscore memb, $\beta$-catenin\textunderscore nuc, Dest\textunderscore compl, E-cadherin, GLI, GSK3$\beta$, MEK, NOTCH, SMAD, SNAI1, SNAI2, SOS/GRB2, TGF$\beta$R, TWIST1, ZEB1, ZEB2, miR200.

The average degree of the T-LGL leukemia network is 1.43, the standard deviation is 0.65 for the in-degree and 0.94 for the out-degree. The average degree of the EMT network is 3.68, the standard deviation is 1.80 for the in-degree and 2.38 for the out-degree. To explore the relationship between the biological case studies and random ensembles, we randomize the biological networks to form an ensemble of 1000 networks for each. We consider two types of randomization, one that preserves the degree of each node (DPR) and one that additionally preserves the regulatory function of each node (DFPR). To preserve node degree, for each randomization, we exchange the child nodes of two randomly selected edges for $50\times M$ times, where M is the total number of edges. Then we generate effective Boolean functions with the same bias as the original network (in DPR) or we just keep the original function (in DFPR). 


As shown in Table \ref{table:DP}, the average damage probability of the two steady states of the  T-LGL network is 0.667 for single node damage and 0.871 for double node damage. The average damage probability of the two steady states of the EMT network is lower, 0.316 for single node damage and 0.449 for double node damage. Table \ref{table:DP} also indicates that after either single node damage or double node damage, the damage probability of the T-LGL leukemia network does not statistically deviate from the randomized ensemble average. The damage probability of the EMT network deviates from the  damage probability of the degree preserving randomized ensemble. The main contributing factor is that the Boolean functions of the EMT network are all canalizing functions and a majority of them are also nested canalizing, which tends to decrease the damage probability, as discussed in the last paragraph of Sec. \ref{sec:III.A.1}. This nested canalizing nature is destroyed when random functions are used, even if they have the same bias. Indeed, the damage probability of the DFPR ensemble is much closer to the result of the EMT network. Though the Boolean functions used in T-LGL leukemia networks are also mostly nested canalizing, the T-LGL network has an average node degree smaller than 2 and most two-input effective functions are also nested canalizing functions, so the randomized functions would be similarly canalizing.


As shown in Table \ref{table:DD}, the probability distribution of each class of double node knockout for the T-LGL network is close to the ensemble average in that class 6 and class 4 are the most well represented. In contrast, the probability distribution of the double node knockout classes for the EMT network is different from the ensemble average in that class 1 is more represented and class 6 is less represented, which is consistent with the deviation in damage probability between the EMT network and its randomized ensemble. In both cases, the result of the biological network is closer to the result of the DFPR ensemble than the result of the DPR ensemble, as expected. We include the detailed classification of double knockout pairs in the two biological networks in Appendix \ref{appendix:2} .

We also apply our algorithm of stabilizing two steady states simultaneously after a single node damage to both networks.  While one generally wishes to eliminate rather than repair a disease state in a biological network, these networks nonetheless provide a useful framework for applying our methodology; after considering joint repair, we will turn our attention to removing the disease state. Most of the nodes have opposite states in the two steady states of both networks. This is not surprising since the two steady states correspond to two opposite biological outcomes  in each case (apoptosis versus survival in the T-LGL network; epithelial versus mesenchymal state in the EMT network). Furthermore, the network reduction used in both cases eliminates nodes that are fixed by source nodes and have the same state in both steady states \cite{ref32,ref33}.

\begin{table}[!tbp]
\squeezeup \squeezeup\squeezeup
\centering
\caption{Comparison between the damage probability of the two biological networks and their randomizations}
\label{table:DP}
\begin{tabular}{|l|l|l|l|}
\hline
network$\backslash$ensemble & DP    & \begin{tabular}[c]{@{}l@{}}DPFR\\ DP,\quad  std\end{tabular} & \begin{tabular}[c]{@{}l@{}}DPR\\ DP,\quad  std\end{tabular} \\ \hline
T-LGL SD         & 0.667 & 0.674, 0.115                              & 0.725, 0.135                                          \\ \hline
T-LGL DD         & 0.871 & 0.875, 0.078                              & 0.905, 0.092                                          \\ \hline
EMT\quad  SD           & 0.316 & 0.449, 0.122                      & 0.792, 0.096                                          \\ \hline
EMT\quad  DD           & 0.473 & 0.689, 0.122                      & 0.943, 0.041                                          \\ \hline
\end{tabular}
\caption*{The $2^{nd}$ column is the damage probability (DP) of the T-LGL leukemia network (first two rows) and the EMT network (last two rows) after single node damage (SD) or double node damage (DD). The $3^{rd}$ and $4^{th}$ columns report the average damage probability and its standard deviation (std) among networks for the corresponding randomized network ensembles. \squeezeup \squeezeup\squeezeup}
\end{table}

The only two nodes having the same state in the T-LGL leukemia network are CREB and IFNG, which exist in a sink branch of the network and do not directly determine the cell state. Thus when we consider the simultaneous repair of the two steady states, there will be 12 cases wherein the damaged node is ON in one state and OFF in the other.   Among them, 4 cases (Caspase, FLIP, IAP, or SMAD knockout) fall into class 7 (see Table \ref{table:3}). Directly damaging the node Caspase may be not biologically interesting as we treat this node to be the sink node of the signaling network here. All the other 8 situations (BID, Ceramide, DISC, Fas, GPCR, MCL1, S1P, sFas) fall into class 8.  The class distribution of the classifications of the DFPR ensemble concentrates on class 8, class 7 and class 4, and the result of the DPR ensemble concentrates on class 8, class 4 and class 7, both in decreasing probability.  The class distribution of the randomized ensembles is consistent with, but with higher spread than the class distribution  of T-LGL leukemia network, which is restricted to class 8 and class 7. The latter is mainly due to the fact that  the two steady states are almost exactly opposite.

When stabilizing the two steady states of the EMT network simultaneously after a single node damage, there are 17 nodes whose knockout can be considered (Dest\textunderscore compl and SOS/GRB2 are OFF in both steady states). Among them, two cases (AXIN2, SNAI2 knockout) belong to class 1 (see Table \ref{table:3}); nine cases (AKT, $\beta$-catenin\textunderscore nuc, GLI, NOTCH, SMAD, TGF$\beta$R, TWIST1, ZEB2, miR2000) belong to class 7 ; six cases ($\beta$-catenin\textunderscore memb, E-cadherin, GSK3$\beta$, MEK, SNAI1, ZEB1) belong to class 8.  As an example, let us consider permanently knocking out node GSK3$\beta$, which is ON in the epithelial steady state and OFF in the mesenchymal steady state. One needs to repair node AKT, MEK, SNAI1, NOTCH and there are 14, 14, 10, 14 simple repair choices for each corresponding node (see Appendix \ref{appendix:3}). As most of the nodes have opposite states in the two steady states, the majority of the repair solutions will be compatible with the other steady state. The algorithm then calculates that there will be 11, 11, 6, 11 repair choices for each corresponding node. The specific choices are listed in Appendix \ref{appendix:3}.

The class distribution of the DFPR ensemble concentrates on class 1, class 8 and class 4 and the result of the DPR ensemble concentrates on class 8, class 5 and class 4, both in decreasing probability. The distributions of the randomized ensembles are more uniformly spread compared with the distribution of the classifications of the EMT network, which is restricted to class 8, class 7 and class 1. This restriction is largely due to the fact that the two steady states are almost exactly opposite.

\begin{table}[!tbp]
\squeezeup \squeezeup\squeezeup
\centering
\caption{Class probability distribution after double node knockout in the two biological networks}
\label{table:DD}
\begin{tabular}{|l|l|l|l|}
\hline
Ensemble $\backslash$ Class & 1     & 4     & 6     \\ \hline
T-LGL          & 0.097 & 0.484 & 0.387 \\ \hline
T-LGL DFPR     & 0.095 & 0.444 & 0.420 \\ \hline
T-LGL DPR      & 0.076 & 0.397 & 0.495 \\ \hline
EMT            & 0.505 & 0.398 & 0.065 \\ \hline
EMT DFPR       & 0.274 & 0.479 & 0.202 \\ \hline
EMT DPR        & 0.041 & 0.326 & 0.609 \\ \hline
\end{tabular}
\caption*{The first and fourth row show the results of the real biological networks, while the rest are the results of their DFPR and DPR ensembles. We only list the three classes (columns) from Table \ref{table:1} that have the highest probability. The probability of the rest of the classes is smaller than 0.04 and does not show much difference between the real network and its randomization.\squeezeup\squeezeup\squeezeup\squeezeup\squeezeup}
\end{table}

In summary, we compared the biological networks with their randomized ensembles, focusing on their damage probability, class distribution after double node knockout and class distribution of repairing two steady states after single node knockout. The T-LGL leukemia network agrees with its randomized ensembles and the EMT network deviates from its degree-preserving randomized ensemble in all tested aspects.


In both networks, a more biologically meaningful intervention than preserving both steady states is to keep the normal steady state as intact as possible and destabilize the disease steady state. If the node to be repaired has opposite states in the two steady states, adding a new edge starting from a node that has the same state in the two steady states will destabilize the disease state.

For example, if Fas is knocked out in the T-LGL network, we need to repair Ceramide to be ON to avert cascading damage to the healthy steady state. The algorithm will give 9 repair solutions involving a new independent edge, shown in Appendix \ref{appendix:1}. Two edge repair solutions (``Ceramide= $\cdots$ OR NOT IFNG'' and ``Ceramide= $\cdots$ OR NOT CREB'', where $\cdots$ stands for the original rule for Ceramide) are not compatible with the second steady state since these two nodes have the same node state in the two steady states. Thus either of these two repair strategies will make the disease state a transient state and the system will keep evolving. Whether the system evolves toward the healthy steady state depends on the node knocked out and the repair solution. In the example above, if we knock out Fas and fix Ceramide to be ON by adding an edge from CREB, this will make the system evolve towards another steady state wherein Caspase is ON, a state biologically similar to the healthy steady state.

Similarly to the T-LGL leukemia network, a repair solution using nodes with the same node state in the two steady states can preserve the epithelial steady state and perturb the mesenchymal steady state of the EMT network. However, in some cases the new attractor is not an epithelial one (E-cadherin is not guaranteed to be ON). 
\squeezeup\squeezeup\squeezeup
\section{Discussion and Conclusion}
\squeezeup\squeezeup
One promising approach to mitigating the effects of diseases is to proactively manipulate the interactions in the relevant biological network. For example, cancerous cells fail to undergo natural cell death; compensatory interactions in the cancer signaling network may in principle drive cancerous cells to undergo cell death. While a theoretical basis for such manipulation has been established in the case of deregulation of a single node (e.g. a single genetic mutation) \cite{ref1}, complex diseases are triggered by several co-existing gene mutations \cite{ref10, ref14, ref15}. The algorithm presented here can be used to design preventive interventions for combinations of multiple dysfunctions of the network. Our identified repair strategy classes provide a framework to explore the short-term combinatorial effects of double knockouts and can be straightforwardly adapted to other types of multiple perturbations. 

The network ensembles most considered here exist in the chaotic phase for very large networks according to the well-studied annealed approximation (due to the average in-degree of 2 or 3), where the topology and update functions are randomized after each time step \cite{ref24,ref30,ref29,ref28}. Thus, we expect the effects of network perturbations to propagate throughout the network. However, to gain detailed insight into the dynamic behavior of the network and to determine specific repair strategies, it is necessary to consider a fixed network topology and interaction rules. We therefore consider two specific biological case studies in this report. 

As patients are often diagnosed with complex diseases after symptoms already developed, the cascading effect of the initial gene mutation or protein dysfunction is already in progress. Thus it is interesting to consider the long-term effects of damage when aiming to repair the effects of single or multiple dysregulations. One can define a node$'$s region of influence as the nodes whose states will be changed due to the cascading effect of its perturbation. Similar to what we have done in the short-term setting, if the regions of influence of two nodes do not intersect and are not co-regulating another target, then the two damage processes are independent of each other, and one would expect to be able to mitigate their effects independently. If the regions of influence of two initially damaged nodes intersect or co-regulate a third node, combinatorial effects will appear and can be analyzed in a similar way as we did here.

In some cases two or more steady states with distinct biological meanings, such as natural cell death and cancerous persistence, may exist \cite{ref10, ref35}. As demonstrated in two biological case studies, our algorithm provides strategies to find compatible ways to stabilize two steady states or stabilize one and destabilize the other. The approach we take here is most useful in designing preventive interventions for disease, as the repair is assumed to be effective on a faster timescale than the propagation of damage. Model-based design of therapeutic methods for complex diseases entails an understanding of the disease state and the identification of manipulations that drive the system from the disease state back to a normal state \cite{ref36}. As a first step, our method provides choices to destabilize the disease state and a framework to test the feasibility of simple edge modifications. A systemic study of the trajectories from a destabilized disease state into a normal state would be another interesting area for future work.

\squeezeup\squeezeup\squeezeup\squeezeup\squeezeup
\appendix
\section{Additional simplification of T-LGL leukemia network}
\label{appendix:0}
\squeezeup\squeezeup\squeezeup
When the sink node Apoptosis is activated, the cell is going to die. Zhang \textit{et al.} chose to represent cell death by a state in which Apoptosis is ON and all the other nodes are OFF and implemented it by adding to every node$'$s Boolean function the clause ``AND (NOT Apoptosis)'' \cite{ref31}. Here for simplicity we do not use this additional clause; this is equivalent with considering any steady state that includes Apoptosis=ON as a normal steady state. In the reduced network (Fig. \ref{pic:fig5}) a small motif consisting of TCR and CTLA4 is isolated from the main part of the network. Since the small motif does not influence the apoptotic decision, we ignore it in the analysis. An auxiliary node P2 in \cite{ref31} is removed and we incorporate the effect in the Boolean rule of IFNG. Also, if the cell is already dead, node knockout and constitutive expression have no biological meaning. However, the activation of Apoptosis requires the node Caspase to be ON first. Thus, we delete the node Apoptosis and consider that Caspase is determining the state of the cell.

\squeezeup\squeezeup\squeezeup\squeezeup\squeezeup
\section{Classification of double knockout pairs in the T-LGL leukemia network and EMT network }
\label{appendix:2}
\squeezeup\squeezeup\squeezeup
We apply the algorithm to stabilize a steady state after double node damage to the T-LGL leukemia network. There are five nodes ON in the healthy steady state, and thus there are 10 double knockout cases (see Table \ref{table:4}). Among them, four cases belong to class 4b (see Table \ref{table:1}). Six cases belong to class 6: four in 6b and one each in 6c and 6d . Thus in this example, there are no cases where less repair is needed, and there are two cases where combinatorial effect occurs. For the disease steady state, there are 7 nodes in the ON state and thus 21 double knockout cases. Among them one case belong to class 1, ten cases belong to class 4b and ten cases belong to class 6: three in 6a and seven in 6b. 

We also apply the algorithm to stabilize a steady state after double node damage to the EMT network.  For the epithelial steady state, there are 6 nodes in the ON state and thus there are 15 double knockout cases. Among them, three cases belong to class 1, nine cases belong to class 4b, two cases belong to class 6b and one case to class 6d. For the mesenchymal steady state there are 13 nodes with ON states and thus 78 node pairs. Among them, 44 cases belong to class 1, 1 case belongs to class 2, 2 cases belong to class 3, 28 cases belong to class 4 (one in 4a, twenty-six in 4b, one in 4c), 3 cases belong to class 6 (one in 6a and two in 6b). Thus there are four cases where less repair is needed and two cases where more repair is needed for double knockout compared to the union of two individual single knockouts. 


\begin{table}[!t]

\caption{Classification of double knockout pairs in the T-LGL leukemia network}

\label{table:4}
\begin{tabular}{|l|l|l|l|l|}
\hline
Node pair        & $S_A$                                                & $S_B$       & $S_{AB}$                                                         & class \\ \hline
BID,Caspase      & IAP                                                  & $\emptyset$ & IAP                                                              & 4b    \\ \hline
BID,Ceramide     & IAP                                                  & S1P         & IAP,S1P                                                          & 6b    \\ \hline
BID,DISC         & IAP                                                  & MCL1,FLIP   & \begin{tabular}[c]{@{}l@{}}IAP,Caspase,\\ MCL1,FLIP\end{tabular} & 6c    \\ \hline
BID,Fas          & IAP                                                  & Ceramide    & IAP,Ceramide                                                     & 6b    \\ \hline
Caspase,Ceramide & $\emptyset$                                          & S1P         & S1P                                                              & 4b    \\ \hline
Caspase,DISC     & $\emptyset$                                          & MCL1,FLIP   & MCL1,FLIP                                                        & 4b    \\ \hline
Caspase,Fas      & $\emptyset$                                          & Ceramide    & Ceramide                                                         & 4b    \\ \hline
Ceramide,DISC    & S1P                                                  & MCL1,FLIP   & MCL1,S1P,FLIP                                                    & 6b    \\ \hline
Ceramide,Fas     & S1P                                                  & Ceramide    & S1P,DISC,                                                        & 6d    \\ \hline
DISC,Fas         & \begin{tabular}[c]{@{}l@{}}MCL1,\\ FLIP\end{tabular} & Ceramide    & \begin{tabular}[c]{@{}l@{}}MCL1,FLIP,\\ Ceramide,\end{tabular}   & 6b    \\ \hline
\end{tabular}
\caption*{All ten node pairs (A, B) are list in the first column. The node to be repaired after knocking out A, B, both A and B are listed in the second to fourth column respectively, where $\emptyset$ means that no node need to be repaired.  The class and subclass index as in Table 1 is listed in the fifth column.\squeezeup\squeezeup\squeezeup\squeezeup\squeezeup}
\end{table}

\squeezeup\squeezeup\squeezeup\squeezeup
\section{Simple solutions after knocking out GSK3$\beta\ $in the healthy steady state of EMT network}
\label{appendix:3}
\squeezeup\squeezeup\squeezeup
All the solutions have similar format as above, where $\cdots$ stands for the original rule for that node. Solutions in normal text format are simple solutions compatible with disease steady state after knockout GSK3$\beta\ $in the healthy steady state, i.e., solutions in \textit{italics} are simple solutions incompatible with disease steady state after knockout GSK3$\beta\ $in the healthy steady state.\\ 
Modifications for node AKT:\\
	AKT=$\cdots$ AND GLI\\    
	AKT=$\cdots$ AND MEK\\    
	AKT=$\cdots$ AND NOTCH\\    
	AKT=$\cdots$ AND SNAI1\\    
	AKT=$\cdots$ AND TGF$\beta$R\\    
	AKT=$\cdots$ AND TWIST1\\    
	AKT=$\cdots$ AND ZEB1\\    
	AKT=$\cdots$ AND ZEB2\\    
	AKT=$\cdots$ AND NOT $\beta$-catenin\textunderscore memb\\    
	AKT=$\cdots$ AND NOT E-cadherin\\    
	AKT=$\cdots$ AND NOT miR200 \\    
      $\mathit{AKT=\cdots AND\  Dest\textunderscore compl}$\\  	
      $\mathit{AKT=\cdots AND\ NOT\ AXIN2}$\\    \spaceup
      $\mathit{AKT=\cdots AND\ NOT\ SNAI2}$\\  
Modifications for node MEK:\\
	MEK= $\cdots$ AND AKT\\
	MEK= $\cdots$ AND GLI\\
	MEK= $\cdots$ AND NOTCH\\
	MEK= $\cdots$ AND SMAD\\
	MEK= $\cdots$ AND TGF$\beta$R \\
	MEK= $\cdots$ AND TWIST1\\
	MEK= $\cdots$ AND ZEB1\\
	MEK= $\cdots$ AND ZEB2\\	
	MEK= $\cdots$ AND NOT $\beta$-catenin\textunderscore memb\\
	MEK= $\cdots$ AND NOT E-cadherin\\	
	MEK= $\cdots$ AND NOT miR200\\	
      $\mathit{MEK= \cdots AND\ Dest\textunderscore compl}$\\
      $\mathit{MEK= \cdots AND\ NOT\ AXIN2}$\\	\spaceup
      $\mathit{MEK= \cdots AND\ NOT\ SNAI2}$\\  
Modifications for node SNAI1:\\
	SNAI1= $\cdots$ AND TWIST1\\
	SNAI1= $\cdots$ AND ZEB1\\
	SNAI1= $\cdots$ AND ZEB2\\
	SNAI1= $\cdots$ AND NOT $\beta$-catenin\textunderscore memb\\
	SNAI1= $\cdots$ AND NOT E-cadherin\\
	SNAI1= $\cdots$ AND NOT miR200\\
	$\mathit{SNAI1= \cdots AND\ Dest\textunderscore compl}$\\
	$\mathit{SNAI1= \cdots AND\ SOS/GRB2}$\\
	$\mathit{SNAI1= \cdots AND\ NOT\ AXIN2}$\\  \spaceup
	$\mathit{SNAI1= \cdots AND\ NOT\ SNAI2}$\\  
Modifications for node NOTCH:\\
	NOTCH= $\cdots$ AND AKT\\		
	NOTCH= $\cdots$ AND GLI\\	
	NOTCH= $\cdots$ AND MEK\\	
	NOTCH= $\cdots$ AND SNAI1\\	
	NOTCH= $\cdots$ AND TGF$\beta$R	\\
	NOTCH= $\cdots$ AND TWIST1\\	
	NOTCH= $\cdots$ AND ZEB1\\	
	NOTCH= $\cdots$ AND ZEB2\\	
	NOTCH= $\cdots$ AND NOT $\beta$-catenin\textunderscore memb\\	
	NOTCH= $\cdots$ AND NOT E-cadherin\\	
	NOTCH= $\cdots$ AND NOT miR200\\	
      $\mathit{NOTCH= \cdots AND\ Dest\textunderscore compl}$\\	
      $\mathit{NOTCH= \cdots AND\ NOT\ AXIN2}$\\	
      $\mathit{NOTCH= \cdots AND\ NOT\ SNAI2}$\\	

\squeezeup\squeezeup\squeezeup\squeezeup\squeezeup\squeezeup\squeezeup\squeezeup\squeezeup\squeezeup\squeezeup
\section{Modifications for node Ceramide after knockout of Fas in the T-LGL leukemia network:}
\label{appendix:1}
\squeezeup\squeezeup
All the solution have the same format: ``Ceramide = $\cdots$ OR New Rule'', where $\cdots$ stands for the original rule. \\
	  Ceramide=$\cdots$ OR BID         \\		 
	  Ceramide=$\cdots$ OR Caspase	\\	 
	  Ceramide=$\cdots$ OR DISC		 \\
	  Ceramide=$\cdots$ OR NOT FLIP	\\ 
	  Ceramide=$\cdots$ OR NOT GPCR\\		 
	  Ceramide=$\cdots$ OR NOT IAP   \\		 
	  Ceramide=$\cdots$ OR NOT MCL1\\		 
	  Ceramide=$\cdots$ OR NOT SMAD\\	 
	  Ceramide=$\cdots$ OR NOT sFas  \\		

\squeezeup\squeezeup\squeezeup\squeezeup\squeezeup\squeezeup\squeezeup
\begin{acknowledgments}
\squeezeup\squeezeup\squeezeup
 This work was supported by NSF Grants PHY-1205840 and IIS-1161007. The authors thank Jorge G. T. Za$\mathrm{\tilde{n}}$udo for useful discussions. We thank two anonymous reviewers for helpful suggestions and for letting us know about reference \cite{syntheticlethality3}.
\squeezeup\squeezeup\squeezeup
\end{acknowledgments}

\squeezeup\squeezeup\squeezeup
\bibliography{network}

\end{document}